%% file: main.tex
\documentclass[a4paper,twocolumn,11pt,unpublished]{quantumarticle}
\pdfoutput=1
\usepackage[utf8]{inputenc}
\usepackage[english]{babel}
\usepackage[T1]{fontenc}
\usepackage{hyperref}
\usepackage[numbers]{natbib}
\usepackage{aas_macros}
\usepackage{amsmath}
\usepackage{amssymb}
\usepackage{url}
\usepackage[nonumberlist, acronym, toc]{glossaries}

\usepackage{braket, mathtools, bm, siunitx, tcolorbox}

\begin{document}

\include{glossary}

\title{Experimental differentiation and extremization with analog quantum circuits}

\author[1]{Evan Philip}
\orcid{0000-0001-5405-2306}
\author[1]{Julius de Hond}
\orcid{0000-0003-2217-934X}
\author[1]{Vytautas Abramavicius}
\orcid{0000-0003-3154-3289}
\author[1]{Kaonan Micadei}
\orcid{}
\author[1]{Mario Dagrada}
\orcid{0000-0001-8825-9049}
\author[1]{Panagiotis Barkoutsos}
\orcid{0000-0001-9428-913X}
\author[1]{Mourad Beji}
\orcid{}
\author[1]{Louis-Paul Henry}
\orcid{0000-0003-1763-2621}
\author[1]{Vincent E. Elfving}
\orcid{0000-0002-5105-5664}
\author[1]{Antonio A. Gentile}
\orcid{0000-0002-1763-9746}
\author[1]{Savvas Varsamopoulos}
\orcid{0000-0002-5277-8768}

\affiliation[1]{Pasqal, 7 Rue Leonard de Vinci, 91300 Massy, France}
\maketitle

\begin{abstract}

Solving and optimizing differential equations (DEs) is ubiquitous in both engineering and fundamental science. The promise of quantum architectures to accelerate scientific computing thus naturally involved interest towards how efficiently quantum algorithms can solve DEs. 
Differentiable quantum circuits (DQC) offer a viable route to compute DE solutions using a variational approach amenable to existing quantum computers, by producing a machine-learnable surrogate of the solution. 
Quantum extremal learning (QEL) complements such approach by finding extreme points in the output of learnable models of unknown (implicit) functions, offering a powerful tool to bypass a full DE solution, in cases where the crux consists in retrieving solution extrema. 
In this work, we provide the results from the first experimental demonstration of both DQC and QEL, displaying their performance on a synthetic usecase. 
Whilst both DQC and QEL are expected to require digital quantum hardware, we successfully challenge this assumption by running a closed-loop instance on a commercial analog quantum computer, based upon neutral atom technology. 
\end{abstract}

\section{Introduction}
\label{sec:intro}

Solving \glspl{de} is a cornerstone task in scientific and engineering disciplines, underpinning models in physics, chemistry, climate modeling, and quantitative finance. While classical numerical methods such as scale-resolving (e.g. finite element, finite volume methods) and spectral solvers~\cite{leveque2007finite,boyd2013chebyshev} remain powerful, their applicability is often challenged by high-dimensional domains, nonlinearities, or stiff dynamics~\cite{Gear1981,Knio2006,Dolgov2020}. 
These limitations have inspired the rise of \gls{sciml}, where trainable models incorporate domain knowledge, enabling more flexible, generalizable and data-adaptive solvers~\cite{rackauckas2020universal}. 
A prominent candidate in this respect have been \glspl{pinn}~\cite{Raissi2019}, tentatively applied also to fluid dynamics~\cite{Cai2021pinn-review}, i.e. traditional domain of scale-resolving methods.  
Despite this promise, \glspl{pinn} can suffer from high training overheads and optimization instability, particularly when scaling to large or complex systems~\cite{Yang2021bpinns, Markidis2021pinns}.

Quantum computing (QC) has emerged as a potential accelerator for scientific computing tasks, e.g. due to its efficiency in solving linear algebra problems and its compact representation of high-dimensional spaces~\cite{harrow2009quantum,Montanaro2016}. While protocols such as HHL and quantum signal processing offer provable advantages for linear \glspl{de}~\cite{childs2017quantum,Lin2020optimalpolynomial}, their reliance on amplitude encoding, full state readout, and large-scale error-corrected devices renders them impractical for current hardware. Nonlinear differential equations bear additional challenges in this framework~\cite{tennie2025quantum}, addressed by hybrid approaches or additional pre-processing - e.g. linearization techniques ~\cite{JPLiu2021,Krovi2023improvedquantum, farghadan2025fast} — which might fall short at capturing strong nonlinearities.
On the contrary, variational quantum protocols cannot rely on provable advantage, but can natively address non-linearities~\cite{lubasch2020variational}, as well as building on classical SciML ideas, like the \gls{dqc} protocol~\cite{dqc}. 
The latter ones adopt the output of parameterized quantum circuits in place of classical neural networks to represent the solution to a differential equation, thus inheriting a seamless integration of data and generalisation properties. 
The potential of \gls{dqc} has been presented in a number of showcases, ranging from fluid-dynamics~\cite{ghosh2022harmonic} and weather modelling~\cite{Jaderberg2024weather}, to extensions involving generative models~\cite{Kyriienko2024protocols}.
A crucial requirement for this strategy is the ability to compute gradients of quantum observables with respect to circuit parameters—a task accomplished by quantum-compatible techniques such as the \gls{psr}~\cite{mitarai2018quantum,kyriienko2021generalized}. 
Also the \emph{readout problem} - affecting quantum algorithms and often a crux when estimating their advantage against classical counterparts~\cite{cotler2021revisiting}- has been investigated in \gls{dqc}-inspired algorithms by dedicated research~\cite{Williams2024readout}.
In certain scenarios like design optimization, the interest in solving a \gls{de} mainly consists in identifying extremal point of the sought solution~\cite{martins2021design}. 
Quantum \gls{sciml} approaches have also been tailored to this specific scenario: e.g. \gls{qel} bypasses the explicit form of the function describing a problem's solutions, and focuses on finding the feature value which extremizes the output of a learned model for such (implicit) function, whilst allowing as input either discrete or continuous data~\cite{qel}.

While variational approaches are theoretically promising and allow for near-term implementation, few experimental demonstrations exist that validate their feasibility to solve \glspl{de} on quantum hardware~\cite{pool2024nonlinear, schillo2025expqcl}.
In particular, neither DQC nor QEL has ever been experimentally implemented on a quantum computer. 
\gls{naqpu} use individual atoms (for instance Rubidium) as qubits, trapped and manipulated via laser-based optical tweezers in vacuum chambers. These platforms offer flexible qubit topologies by allowing atoms to be arranged in custom spatial patterns. 
Depending on the atomic states used, \gls{naqpu} support multiple computational modes. Beside the well-known universal, \emph{digital} quantum computing paradigm recently experimentally demonstrated~\cite{bluvstein2024logical, reichardt2024logical}, \emph{analog} computing using Rydberg states where atom interactions implement e.g. a quantum Ising model is also a viable opportunity to implement quantum algorithms~\cite{henriet2020quantum}.
Additionally, hybrid digital-analog approaches exist, that combine aspects of both offering additional schemes tailored to specific tasks.
Crucially, even if variational quantum algorithms are often discussed in the quantum circuit model, that is, in terms of gates acting on non-interacting qubits, there is no fundamental barrier against implementing them on a digital-analog quantum computer~\cite{daqc}.

Inspired by the latter consideration, the goal of this work is to present a closed-loop example of both \gls{dqc} \& \gls{qel} on a cloud-available quantum computer offering native analog operations, adopting ad-hoc differentiation rules for the quantum circuits. 
The architecture of choice is a \gls{naqpu} utilizing individual $\prescript{87}{}{\text{Rb}}$ atoms trapped in an array of optical tweezers that operates in the ground-Rydberg qubit basis with global analog control~\cite{fresnel}.
Several details of the algorithmic protocols, for instance the construction of an appropriate \gls{fm} and the differentiation of the circuit against the input variables,  required ad-hoc adaptations that are presented alongside with the results of a complete example solving and maximising a \gls{de} invoking quantum hardware. 
Therefore, our paper represents a first step and a useful practical guide on how to operate \gls{de}-processing variational algorithms in a digital-analog framework.

\section{Methodology} 
\label{sec:methodology}

We here focus on illustrating the experiment design to validate \gls{dqc} \& \gls{qel} algorithms in our platform. For clarity, we limit our discussion to cases involving a single independent variable $x$ to match the exemplary run on quantum hardware reported here, though it is certainly possible to generalise to multiple variables.

The proposal to solve \glspl{de} with a quantum circuit \cite{dqc} rests upon the following idea: given e.g. for simplicity an ordinary \gls{de} $\frac{df}{dx}=g(f(x), x)$ in a single variable $x$, we can tune the output  $f_{\boldsymbol{\theta}}(x)$  of a quantum circuit defined by a set of parameters $\boldsymbol{\theta}$ 
to approximate the solution ${f}(x)$. 
The section of a quantum circuit introducing such parameterisation is often called the \emph{ansatz}, and labelled as a unitary evolution  $U_\text{a}(\boldsymbol{\theta})$. Various architectural choices are possible for $U_\text{a}(\boldsymbol{\theta})$: see \cite{bharti2022noisy} for a succinct list. 
Inheriting from \gls{pinn} methodology, variationally approximating the solution means to minimise against $\boldsymbol{\theta}$ a formal \emph{loss function} 
\begin{equation}
\label{eq:loss}
L_d(\boldsymbol{\theta}) =\sum_{\{x_i\}} 
\left( \frac{df_{\boldsymbol{\theta}(x)}}{dx} - g(f_{\boldsymbol{\theta}}(x), x)
\right)^2 dx,     
\end{equation}
with $\{x_i\}$ a finite sample of points (due to practicality) within the solution domain. $\{x_i\}$ are called the \emph{collocation points}.
In this regard, the quantum circuit acts as a \emph{universal function approximator}: $f(x)$ and its derivatives $df/dx$ can be estimated by appropriate manipulation and readout of the variational circuit, as detailed in \cite{dqc} and summarised below. 
By construction, $L_d(\boldsymbol{\theta})=0$ when the differential equation holds in any such points.
Boundary conditions can be added as $L_b(\boldsymbol \theta)$ in an analogous way, i.e. as the absolute difference between the known $f(x)$ and its value estimated by the quantum circuit at the chosen boundary point(s). 
Even if exemplified for ordinary \glspl{de}, the same procedure applies equally well to partial \glspl{de}, stochastic \glspl{de} and higher dimensional cases~\cite{dqc, Paine2021}. 

\gls{qel} is a model-based quantum optimization framework designed to identify the input configuration that extremizes (i.e., maximizes or minimizes) a given unknown objective function~\cite{Varsamopoulos2022}. \gls{qel} comprises of two phases: a \textit{learning} phase and an \textit{extremization} phase.
In the learning phase, a model based upon a variational quantum circuit is trained to approximate an unknown target function ${f}(x)$, capturing the relationship between input $x$ and output. This training process for a continuous function is identical to the one employed in \gls{dqc}.
In addition, an extremization phase extends this modelling functionality by incorporating a post-training optimization step. In this stage, the goal shifts to locating the input value $x_{opt}$ that extremizes the trained model's output. 
For continuous input domains, such as those encountered in \glspl{de}, extremization is achieved by holding the trained model parameters $\boldsymbol{\theta}$ fixed and analytically differentiating the trainable/parametrized model with respect to the input ${x}$. Owing to the continuous nature of the inputs, this \gls{fm} is amenable to circuit differentiation techniques, such as the parameter-shift rule. Gradient-based optimization methods, including gradient ascent or descent, can then be applied to iteratively converge to the optimal input $x_{opt}$.

Algorithmically, the aforementioned approaches rest on some key aspects:
(i) encoding the dependency upon all the relevant variables in the quantum circuit -- typically via a section of the circuit itself $U_{\text{f}}({\mathbf{x}})$, i.e. the \emph{\acrlong{fm}};
(ii) being able to vary the output of $f_\theta(x)$ enough to represent the target function -- often loosely referred to as the \emph{expressivity} of the circuit;
(iii) computing efficiently and accurately derivative terms such as  $\frac{df_\theta(x)}{dx}$, which are necessary to estimate the loss $L$ in the case of \gls{dqc} and continuous \gls{qel}.
We focus now on discussing in practice these requirements, as they dictate necessary features of the hardware for a successful implementation of the chosen variational algorithms.

Feature mapping an input variable is a key feature of \gls{qml}, with various strategies having been discussed to make it the most effective, and at the same time the least demanding on the hardware, e.g.~\cite{bharti2022noisy}. 
At a minimum, we necessitate a nontrivial dependence of the output on the input variable $x$, and the ability to vary this dependence reliably. 
For algorithms like \gls{dqc}, which invoke a \gls{nisq}-friendly angle-embedding~\cite{dqc}, this translates in the output being a convolution of multiple, distinct contributions $e^{-i \omega_j \phi(x)}$, where the function $\phi$ encodes the input variable. 
Notably, the frequency spectrum $\{\omega_j\}$ is determined \emph{exclusively} by the \gls{fm} (inclusive of ad-hoc classically tunable hyperparameters~\cite{Jaderberg2024}).
Within the {ansatz}, altering the set of tunable parameters $\boldsymbol{\theta}$ changes the output and derivatives circuit with respect to the input variable, as it mixes in different proportions the frequency components $\{\omega_j\}$ set by the \gls{fm}.

Optimization loops typically imply the ability to compute derivatives with respect to the ansatz parameter(s) $\boldsymbol{\theta}$. 
Also, to perform optimization loops against loss functions which contain a $df/dx$ derivative like Eq.~\eqref{eq:loss}, we need to be able to estimate this fundamental term in the workflow.
Both DQC and QEL rely on the aforementioned \gls{psr} to calculate derivatives instead of finite difference methods \cite{psr1, psr2}, which are impractical due to experimental noise, beside the fundamental limitations already exposed in the introduction. 
\gls{psr} can extract analytically derivatives $\frac{d^nf}{d\theta_i^n}$ of order $n$ against a certain parameter in the circuit $\theta_i$~\cite{mitarai2018quantum, Schuld2019parametershift}.
This idea rests on the fact that we can interpret the unitary evolution dictating the quantum circuit output $f(x, {\theta_i})$ as a smooth transformation $M_{\theta_i}$ for each parameter. For example, in the case of a single gate we would have $M_{\theta_i}(\cdot ) \equiv U^\dagger (\theta_i) (\cdot) U(\theta_i) $, where $(\cdot)$ is typically a cost operator~\cite{mitarai2018quantum, dqc}.  
A first order derivative $\frac{df}{d\theta_i}$ can then be expressed in many cases of interest by executing two equivalent circuits, where the target parameter of the transformation is displaced by an appropriate amount -- e.g. $M_{\theta_i\pm \pi/2}$, and linearly combining their output in appropriate fashion.
Higher order derivatives can then be computed by iterating such rule and additional, more complex cases can rely on the product and chain rules of derivation to be reduced to this fundamental update rule. 
Differentiating the \gls{fm}, i.e. computing $\frac{d^nf}{dx^n}$, proceeds in a similar fashion, where the terms involving the shifted unitaries depend crucially on the unique spectral gaps characterizing the \gls{fm}~\cite{dqc, Jaderberg2024}.   

However, \gls{psr} is not applicable in the chosen setting of analog \glspl{naqpu}. The always-on interactions present in the latter architecture violate some assumptions required to implement \gls{psr}, i.e that the action of the target parameter is described by an involutory matrix~\cite{Schuld2019parametershift}. 
Generalized versions for analytic differentiation exist to tackle such cases~\cite{gpsr}, but they are resource intensive. 
An experimentally viable option is to adopt gradient-free strategies, like sampling and interpolating the output function. 
However, since analytic derivatives and derivative circuits play a central role in \gls{dqc}, here we demonstrate instead the recently developed \gls{agpsr}~\cite{agpsr-publication} in order to study its feasibility in an experimental setting. 
For more details on (feature) circuit differentiation, we refer the reader to Appendix \ref{appsec:differentiation}.

\begin{figure*}[!htb]
    \centering
    \includegraphics[width=\linewidth]{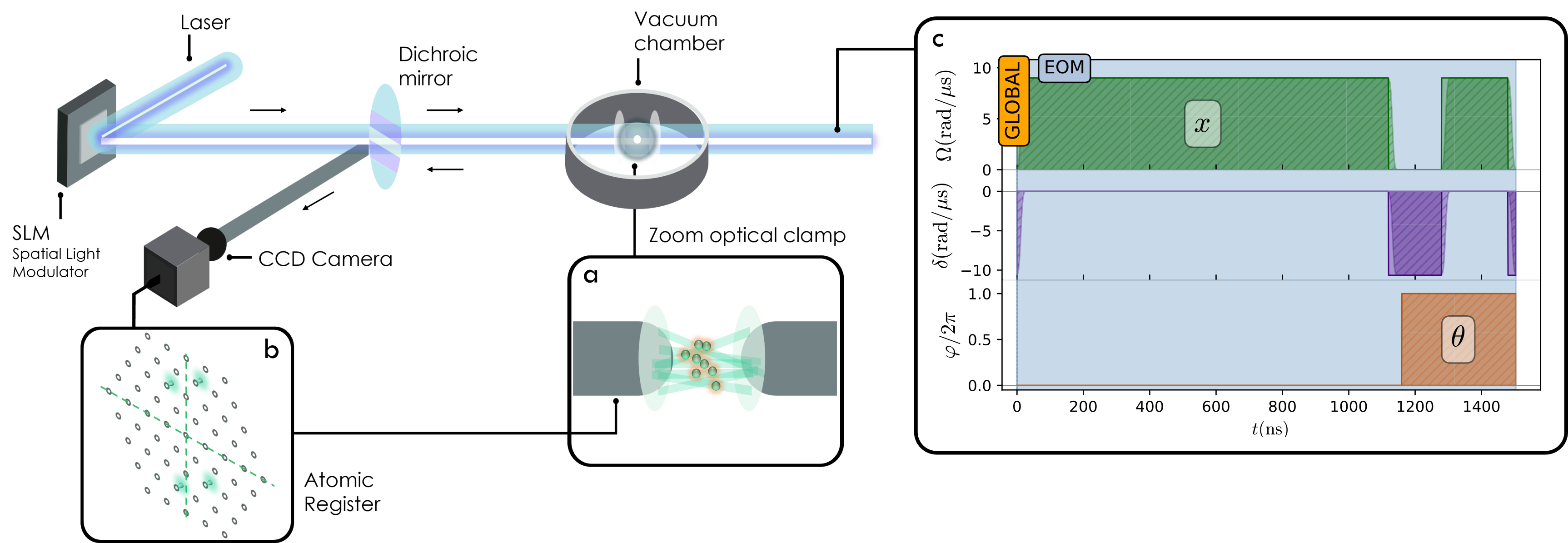}
    \caption{\emph{The neutral-atom register.} A pictorial representation of the setup used for the experiment, along with an indication of the main components. 
    \textbf{a} A zoom on the optical clamping of the \textit{Rb} atoms, attained via the tweezers inside the vacuum chamber. 
    \textbf{b} Perspective representation of the regular hexagonal 61-atoms array generated for each circuit execution, with the two multiplexed qubit sets targeted for the experiment highlighted in green.
    \textbf{c} \emph{Feature Map and trainable ansatz via laser pulse sequence.} The area of the first pulse $\Omega$ maps the value of the feature variable $x$, whereas the phase $\phi$ of the last pulse corresponds to the ansatz parameter $\theta$. 
    The modulation of the ideal square pulses (as solid lines) is visible as overlapped shaded areas.
    This inset was generated using \texttt{pulser}~\cite{pulser} and shows the longest sequence, with the highest phase value that was used in this work. 
    \label{fig:register}}
\end{figure*}

\subsection{Problem identification}
\label{sec:problem}

\begin{figure*}[!htb]
    \centering \includegraphics{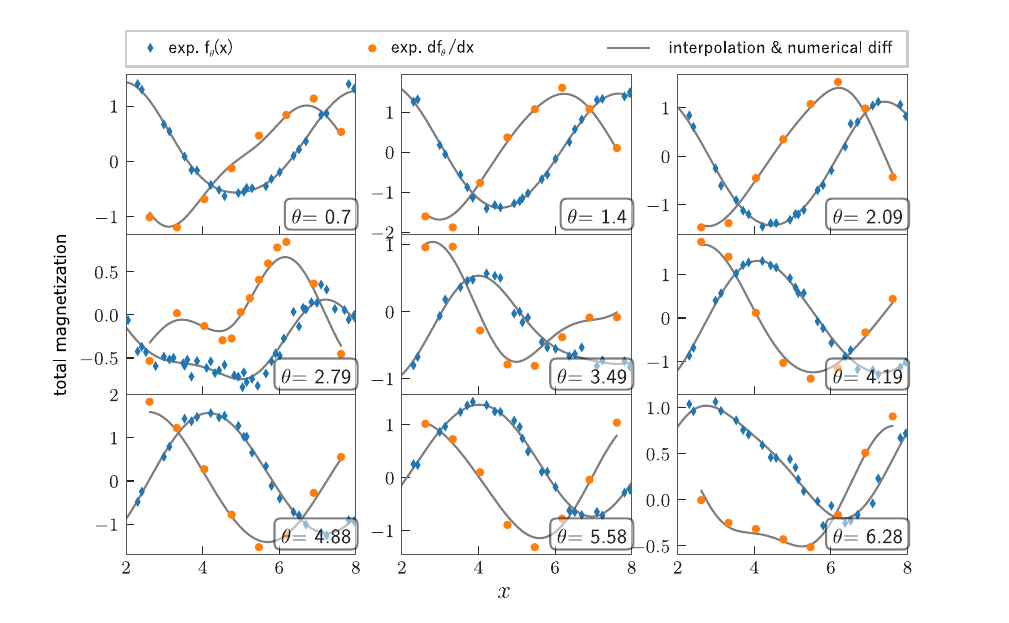}
    \caption{\emph{Derivatives} calculated after applying the \gls{agpsr} (orange dots) to the output of the the parameterised quantum circuit of Fig.~\ref{fig:register}c (blue dots), representing $f_\theta(x)$. For comparison, we report alongside the function and its derivative, as calculated from the interpolated smoothed data (grey lines). 
    We present the results for each of the 9 values of the ansatz parameter $\theta$, as elicited in the legend of each plot. 
    The x-axis represents the value of the input variable $x$.  
    \label{fig:derivative}}
\end{figure*}

We choose a problem apt for embedding in the current hardware and analytically solvable to have a rigorous benchmark for the solution points computed with the quantum device.
At the same time, we ensure to test all primitives and stages of a combined \gls{dqc} and \gls{qel} process, as discussed above. 
We target the resolution of the first-order, linear initial value differential problem:
    \begin{align}
        &\frac{df}{dx} = \sum_{i= 0, \ldots,  6} \alpha_i x^i  \label{eq:de_rounded}\\
        &f(6.516) = 0, \nonumber
    \end{align}
in the arbitrary domain $x \in (2, 8)$.
The full expression, the chosen coefficients $\{\alpha_i\}$ and arguments behind the choice are in Appendix Eq.~\eqref{eq:de}.
The goal for \gls{dqc} is here to represent accurately $f(x)$ and its derivative in the collocation points chosen within the domain, i.e. to train a surrogate of the \gls{de} solution.
With \gls{qel}, instead, we aim to find the input that minimizes the value of the solution $f(x)$, within the chosen domain, using directly the previously trained \gls{dqc} model. 
In addition to the value of the function at the boundary point $x_b=6.516$, \gls{dqc} requires information regarding the function and its derivative at the collocation points, in order to calculate the loss. 
For this experiment, we adopted $8$ evenly spaced points in the domain, as displayed in Fig.~\ref{fig:derivative}.

To minimise the interaction of the hardware with external control, as further elicited later, we opted for a closed-loop experiment. 
Therefore, we pre-selected 9 evenly spaced points $0.70 \rightarrow 6.28$ for the ansatz single parameter $\theta$, making sure to include a value $\theta_{opt} \simeq 2.79$ where the loss would be minimized, according to simulations.

For the second stage of QEL, after the learning of the model, an optimization is performed with respect to the input $x$. 
Therefore, one requires to estimate $df/dx$ at the optimized ansatz parameter value $\theta_{opt}$. 
To allow for a finer grid near the expected minima, we add at this stage additional 5 points to the collocation set.
Additional details are reported in Appendix \ref{appsec:differentiation}.

\subsection{Experimental setup and design}
\label{sec:setup}

The experiment was performed on an analog \gls{naqpu}~\cite{henriet2020quantum}, accessible via cloud-based queries~\cite{pasqalcloud}. 
In the setup, $^{87}$Rb atoms are trapped in a register of $61$ atoms via optical tweezers generated from laser light using a spatial light modulator (see Fig.~\ref{fig:register}). Each atom encodes a qubit, with the $|0\rangle$ and $|1\rangle$ state represented by the $|5S_{1/2};F=2,m_F=2\rangle$ ground state and $|60S_{1/2};m_J=1/2\rangle$ Rydberg level, respectively. Coupling between these states is implemented with a global analog control channel, which physically consists of two lasers that excite via the intermediate $6P_{3/2}$ state, but which, owing to a large detuning from this state, can be described as a single coupling \cite{Brion07}.
Owing to their large electric polarizability, same-parity Rydberg states have a strong Van der Waals interaction that can be used to create entanglement. The interaction is given by $V=\hbar C_6/r^6$, with $r$ the interatomic distance, while for our Rydberg state of choice $C_6\approx 2\pi \times 138~\mathrm{GHz\,\mu m^6}$.       

The experiment is effectively performed on a sub-register consisting of two qubits, separated by $r\approx 8.7~\mathrm{\mu m}$.
Setting $\hbar = 1$, the Hamiltonian dictating the dynamics of a $n=2$ qubit system in a \gls{naqpu} device is given by:
\begin{widetext}
\begin{equation}
  \frac{\Omega(t)}{2} \left(\cos(\varphi(t)) \sum_{i} \sigma^X_i - \sin(\varphi(t)) \sum_i \sigma^Y_i \right) - \frac{\delta(t)}{2} \sum_i \sigma^Z_i + \frac{C_6}{r^6} N_1 N_2,
\end{equation}
\end{widetext}
where $N_i \coloneqq (1+\sigma^Z_i)/2$ and here $i \in \{1,2\}$ indexes the atom in the array.

In this experiment, the distance $r$ between the qubits is fixed during the execution of a sequence, and together with $C_6$, it sets the interaction strength among the atoms.
The $r$ value was chosen to allow for the interaction term to mix multiple frequencies in the output, as it will appear clearer after Eq.~\eqref{eq:generator}. 
Similar experiments often choose to work in the strongly-blockaded regime instead, such that $C_6/r^6 \gg \Omega$ (see e.g.\ Ref.~\cite{Bernien17}). This bears the advantage of making any experiment less sensitive to fluctuations in the position, but in our case we compromised towards weaker regimes, to allow more flexibility in the ansatz output.

The amplitude, phase and detuning parameters, respectively $\Omega(t)$, $\varphi(t)$ and $\delta(t)$, can be varied as a function of time $t$ during the execution of a sequence modulating the control laser, as exemplified in Fig.~\ref{fig:register}c. They thus represent tunable parameters in the corresponding parameterized quantum circuit. 
The analog configuration of this system only allows for global, identical pulses on the Rydberg channel across the register, as well as limited register configurations, limiting the expressivity of the ansatz. 

A fairly general parameterised quantum circuit for a one-dimensional feature $x$, i.e. the case treated in previous sections, has an output state that can be written:
\begin{align}
    \ket{\psi_\text{o}(x; \boldsymbol{\theta})} &= U_\text{a}(\boldsymbol{\theta}) U_\text{f}(x) \nonumber \\
    &= U_\text{a}(\boldsymbol{\theta}) \exp\left(- i \left(\frac{x}{2}\right)  G_\text{FM}\right) \ket{\varnothing},
    \label{eq:output_state}
\end{align}
where $G_\text{FM}$ represents the generator of the \gls{fm} (as defined in \cite{gpsr}) and $U_\text{a}$ represents the ansatz, tunable via the variational parameters $\boldsymbol{\theta}$, and $\ket{\varnothing}$ is the initial reference state - which for the chosen hardware is approximately $\ket{0}^{\otimes n}$. 
For this experiment, we choose 
\begin{equation}
\label{eq:generator}
    G_\text{FM} =  \left(\frac{2}{\bm{\Omega}}\right) \left( \frac{\bm{\Omega}}{2} \sum_{i=1}^2 \sigma^X_i + \frac{C_6}{r^6} N_1 N_2 \right). 
\end{equation}
such that the \acrlong{fm} term appearing in Eq.~\eqref{eq:circ_output} can be accomplished by subjecting the ground state to a square drive pulse of fixed $\Omega(t) = \bm{\Omega} = \qty{9}{\radian / \micro \second}$, duration $(\frac{x}{2})(\frac{2}{\bm{\Omega}}) = \qty[parse-numbers=false]{\frac{x}{9}}{\micro \second}$, $\delta=0$ and $\varphi=0$ - as shown in Fig.~\ref{fig:register}c.
It is important to notice how the $G_{\text{FM}}$ introduced in Eq.~\eqref{eq:generator} is not involutory or idempotent, and therefore eludes the assumptions behind using \gls{psr} to calculate derivatives with respect to the encoded $x$.  
Therefore, we calculate derivatives via \gls{agpsr} \cite{agpsr-publication}, as mentioned in Section~\ref{sec:methodology}.

In order to perform the training, we estimate via optical measurements the operators $\set{\sigma^Z_1, \sigma^Z_2}$, and combine them in the total magnetization from the state at the circuit output:
\begin{equation}
\label{eq:circ_output}
    \Braket{\psi_\text{o}|\left(\sigma^Z_1 + \sigma^Z_2\right)|\psi_\text{o}},
\end{equation}
which is then used as the cost function~\cite{dqc}.

As aforementioned, this experiment was designed as closed-loop, i.e. no choices were performed as a feedback from the experiment steps. 
The ansatz $U_\text{a}(\theta)$ was accomplished by a second pulse of fixed amplitude $\Omega(t) = \qty{9}{\radian / \micro \second}$, zero detuning $\delta$ and phase $\varphi(t) = \theta$, as shown in Fig.~\ref{fig:register}c.
We performed gradient-free optimization for the single ansatz parameter $\theta$, such that here the exact structure of $U_\text{a}(\theta)$ is irrelevant when considering differentiation rules for the circuit.
The experiment required overall the execution of 308 different sequences (hence not counting repetition to accrue statistical significance from the measurements), as detailed in Appendix \ref{appsec:differentiation}.
In order to accrue statistics from the destructive measurement to estimate Eq.~\eqref{eq:circ_output}, each sequence as in Fig.~\ref{fig:register}c was repeated a number of shots. 

Note that the entire experiment was performed using an electro-optic modulator (EOM) to execute square pulses with a higher modulation bandwidth~\cite{albrecht2023quantum}). As a result, there is a finite detuning during delays, as evident in Fig.~\ref{fig:register}c. This can be considered to be a part of the ansatz and does not affect the demonstration of the algorithm itself.

Finally, to leverage and demonstrate on multiplexing the experiment execution, as made possible by the much larger addressable register of 61 qubits, two copies of the same experiment were run simultaneously. 
With such twofold multiplexing, in effect the number of shots was double the number of executions of the sequence in Fig.~\ref{fig:register}c; nevertheless, the raw data was post-processed to aggregate the shots before further analysis. 
Any unintended interaction between the two copies was minimised by placing them sufficiently far within the register space, i.e. at a distance (between the centers of the two copies) $D \gg (C_6/\Omega)^{1/6}$, see Fig.~\ref{fig:register}.

\section{Results}
\label{sec:results}

As a preliminary step, we performed detailed simulations of the algorithm execution, adopting the \texttt{pulser} open-source emulator~\cite{pulser} for \gls{naqpu} hardware.
The simulation results, using the default hardware configuration, were adopted to calibrate against experimental offsets in the detuning parameter $\delta$, thus aligning the expected total magnetization from ideal simulations with actual values (see Appendix~\ref{appsec:hardware}).

For each of the 308 sequences corresponding to the various estimates required by the algorithm, we used an average of 204.6 shots, without accounting for events such as unsuccessful preparation of the register, resulting in invalid bitstring outcomes.

We attained experimentally the total magnetization value in Eq.~\eqref{eq:circ_output} at 4 shifted values, for each collocation point. 
\gls{agpsr} was applied to calculate the derivative~\cite{agpsr-publication}, with the calculation outcome shown in Fig.~\ref{fig:derivative} for all 9 values of the ansatz parameter $\theta$.
Here, we demonstrate the experimental feasibility of the \gls{dqc} protocol adopting (aG)PSR to compute differentiable circuits, bypassing the shortcomings of numerical differentiation~\cite{dqc}. However for benchmark and comparison, we report also results attained with such standard techniques. In particular, we interpolated  $f_{\theta}(x)$ as estimated in the experiment across the 32 data-points and smoothed it via the Whittaker-Eilers algorithm, to then differentiate numerically the outcome. The interpolated function and corresponding derivative are reported as reference in Fig.~\ref{fig:derivative}, and validate the derivatives computed via \gls{agpsr}. 
    
\begin{figure*}[!htb]
    \centering \includegraphics[width=\textwidth]{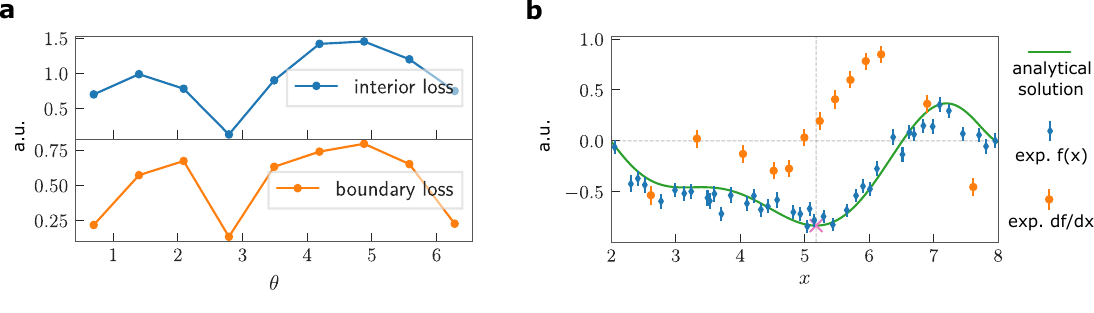}
    \caption{
    \emph{Experimental results for the execution of DQC and QEL protocols on a \gls{naqpu}.}
    \textbf{a} (On the y-axis) the loss as calculated from the $8$ chosen collocation points $\{x_i\}$, (on the x-axis) for various values of the ansatz parameter $\theta$. 
    As blue dots we plot the root of $L_d(\theta)$ in Eq.~\eqref{eq:loss}, whereas as orange dots we plot the boundary loss $L_b(\theta)$. 
    \textbf{b} As blue dots, the total magnetization as obtained from experimental data, representing the target function $f(x)$ as estimated via the parameterised quantum circuit, with the pink cross marking its minimum. 
    For comparison, the green line shows the analytical solution of the differential equation, Eq.~\eqref{eq:de_rounded}.
    As orange dots, the derivative $df/dx$ calculated by \gls{agpsr}.
    Error bars for the function (derivative) estimates represent the (propagated) shot noise from the device. 
    On the x-axis we report the value of the input variable $x$. 
    The gray dashed lines mark the analytical minimum location, and the zero of the y-axis. 
    \label{fig:results}}
\end{figure*}

The experimentally attained predictions of $f_{\theta}(x)$ and its derivative $df_{\theta}/dx$ , for each collocation point and all the chosen values of $\theta$, were used to estimate the loss function in Eq.~\eqref{eq:loss}, and the outcomes are reported in Fig.~\ref{fig:results}a. 
The value of $\theta$ corresponding to the minimal loss ($\theta_{\text{opt}}=2.79$) was selected as the best candidate for the solution of the \gls{de}, in a manner similar to grid search black-box optimization. 
The solution obtained for $f(x)$ along with its derivative are shown in Fig.~\ref{fig:results}b, along with the analytical solution.
The close match between the latter and experimental data flags the successful execution of the \gls{dqc} phase of the experiment, as the quantum circuit was able to solve the targeted \gls{de} in Eq.~\eqref{eq:de_rounded}.

We can now move to observe whether, fixing the learned value $\theta = 2.79$, we can implement the second stage of the experiment, i.e. the \gls{qel} protocol. 
The gray lines represent the smoothed interpolation of the data and the derivative of this interpolation.
This consists in obtaining the value of the $x$ input variable, that minimizes $f(x)$. 
In order to perform an accurate estimate, the derivative w.r.t. $x$ was obtained at 5 additional collocation points, as it can be observed comparing the subplots in Fig.~\ref{fig:derivative}. 
At the value of parameter $\theta$ corresponding to the trained function, these 13 derivatives $df/dx$ were used to determine the value of $x$ that minimizes $f(x)$, thus emulating a gradient-based search algorithm. 
It is evident from Fig.~\ref{fig:results}b how the derivative approaches zero multiple times, corresponding to the inflection point and maxima in the exact solution, which is expected. 
Among such candidates, the value of $f(x)$ known from the quantum circuit estimates can be easily adopted to rule out false-positives, and identifying as the learned extremal point $x_{\text{opt}}=4.995$, which is reasonably close to the exact minimum as computed analytical from the analytical solution of the \gls{de}, i.e. $\Bar{x}=5.140$.

\section{Conclusions and future work}
\label{sec:conclusion}

\glsresetall

With this work, we demonstrated the possibility to engineer pulses on a neutral atom NISQ device, to run natively digital protocol to variationally solve a \gls{de}, and extremize them leveraging upon the proposals to use differentiable quantum circuits.
We demonstrated the above on a first-order ordinary \gls{de} defined upon a 1D domain, namely adopting \gls{dqc} and \gls{qel} protocols. Whilst this is a minimal working example, such protocols hold promise to be readily generalized towards more complex examples, as allowable by hardware capabilities.

Whilst the results presented in this work mark a successful completion of a first step towards adopting quantum computers for representing solutions to differential equations, it is immediate to recognize various directions for improvement. 
As the first, the experiment was performed in closed-loop. This did not prevent to demonstrate all the crucial operations of the targeted quantum protocols, nevertheless, an obvious logical step is to check whether the feedback from each training epoch in an open-loop could reduce the number of circuit executions, and hence the overall experiment duration, at the cost of keeping an online connection to the device and ensuring little to no drift in the device characterization.  
In this scenario, the second stage involving QEL, where derivatives w.r.t. the input variable $x$ are queried, would replace the gradient-free grid search adopted here with a gradient descent optimization. 

In this experiment we also highlighted, as a proof-of-principle, the possibility to multiplex such experiments leveraging upon the vast registers available to \glspl{naqpu} hardware implementations~\cite{pichard2405rearrangement, chiu2025continuous}. 
This can be highly beneficial in reducing the required number of shots, especially considering how several analyses of the algorithms displayed here required $\mathcal{O}(10)$ qubits to tackle much more complex benchmark problems with high-quality results. 

From an experimental point of view, the a-posteriori engineering of the ansatz can massively benefit from improvements in the experimental setup of the quantum circuit. Detuning masks, access to hyperfine states and other solutions enabling single-atom addressability with the laser pulses would enable e.g. digital-analog ans{\"a}tze and hence significantly more expressive circuits. For example, one could adopt \emph{tower} \glspl{fm} already disclosed in \cite{dqc}. 
The latter improvements would also enable a systematic comparison between gradient-free against gradient-based methods in realistic experimental conditions, also beyond natively digital hardware. 

In conclusion, our work represents an experimental milestone delivering the first successful differentiation of feature maps attained directly on real quantum hardware, and we believe that the results presented here will inspire conducting further research on variational algorithms on both analog and digital quantum devices. 

\subsection*{Ethics declaration}
Pasqal owns a patent on DQC~\cite{dqc-patent} and QEL~\cite{qel-patent} and has filed a patent for \gls{agpsr}~\cite{agpsr-publication}.

\subsection*{Acknowledgements}
The authors thank for helpful comments and support Pasqal's pulser team (and in particular H. Guimaraes Silverio), as well as the Hardware team for making this experiment possible.
We would also like to thank O. Kyriienko for useful discussions during the preparation of the experiment.

\bibliographystyle{quantum}
\bibliography{biblio}

\onecolumn
\appendix

\newpage

\begin{center}
\section*{Supplementary Materials}    
\end{center}

\section{Details about the differential problem}
\label{appsec:deproblem}

We report here the complete parameterization of the differential equation solved in the DQC stage:
\begin{multline}
\label{eq:de}
    f' = 63.1 -857.7 \left(\frac{x}{8}\right) + 4503.2 \left(\frac{x}{8}\right)^2 - 11823.4 \left(\frac{x}{8}\right)^3 + 16477.2 \left(\frac{x}{8}\right)^4 - 11615.9 \left(\frac{x}{8}\right)^5 + 3253.3 \left(\frac{x}{8}\right)^6 
\end{multline}
We chose an analytically solvable equation to avoid introducing numerical approximations in the benchmark solution.
The choice of the coefficients in the differential equation made sure that the exact solution is in principle expressible by an ideal version of our quantum circuit, upon training.

The following collocation points were chosen for $\{x_i\}$ in the $x$ domain:
\begin{center}
\begin{tabular}{ c c c c c c c c }
2.614 & 3.328 & 4.042 & 4.757 & 5.471 & 6.185 & 6.900 & 7.614
\end{tabular}
\end{center}

The ansatz parameter $\theta$ was varied among the following values: 
\begin{center}
\begin{tabular}{ c c c c c c c c c }
0.70 & 1.40 & 2.09 & 2.79 & 3.49 & 4.19 & 4.88 & 5.58 & 6.28
\end{tabular}
\end{center}

For QEL, the following additional domain points were chosen for calculation of derivatives with respect to the input $x$:
\begin{center}
\begin{tabular}{ c c c c c }
    4.519 & 4.995 & 5.233 & 5.709 & 5.947
    \label{tab:qel_points}
\end{tabular}
\end{center}

\section{Experimental differentiation and extremisation of the circuit}
\label{appsec:differentiation}

We calculate all derivatives, as detailed in the accompanying publication, via \gls{agpsr}~\cite{agpsr-publication}.

After inspecting the generator encoding the input variable $x$ in Eq.~\eqref{eq:output_state}, we chose two shifts for the target variable, i.e. $0.90$ and $2.47$.
Importantly, \gls{agpsr} considers an ideal \emph{square} pulse, using the ideal version of the generator in Eq.~\eqref{eq:generator}. 
Despite the fact that pulses are necessarily smoothened by the generator in the experiment (see for reference the modulation visible in Fig.~\ref{fig:register}c), the derivative attained in the experiment and reported in Fig.~\ref{fig:derivative} matches well the smoothed interpolation on the output data (and ultimately its expected analytical expected values), such that no further correction to the \gls{agpsr} calculations was deemed necessary.

During the training of \gls{dqc}, a total of 8 derivatives $df/dx$ were calculated, stored and interrogated from the classical memory when executing the \gls{qel} stage - without accruing additional calls to the hardware.  
However, as mentioned in the main text, to attain a finer grid near the extremal point, we added 5 additional points in the collocation set (SM \ref{tab:qel_points}), thus attaining a total of 8 evenly spaced points in the restricted interval $[4.519, 6.185]$

For all ansatz parameter values, this resulted in the execution of 32 sequences as in Fig.~\ref{fig:register}c each, resulting from the eight domain points, each shifted four times (the boundary value did not require a separate sequence since it was simultaneously a shifted value). 
The only exception is represented by $\theta=2.79$, where the 5 additional points accrued a total of 52 sequences. 
Therefore, the overall experiment could be executed with only a total of 308 sequences.
Due to heuristic preparation of the register, not all the shots queried were successful, and those aborted were removed from the resulting bitstrings. 

The full raw data is stored in~\cite{datafile} and can be made available upon request.

\section{Hardware configuration and characterization}
\label{appsec:hardware}

\begin{figure*}[!htb]
    \centering 
    \includegraphics{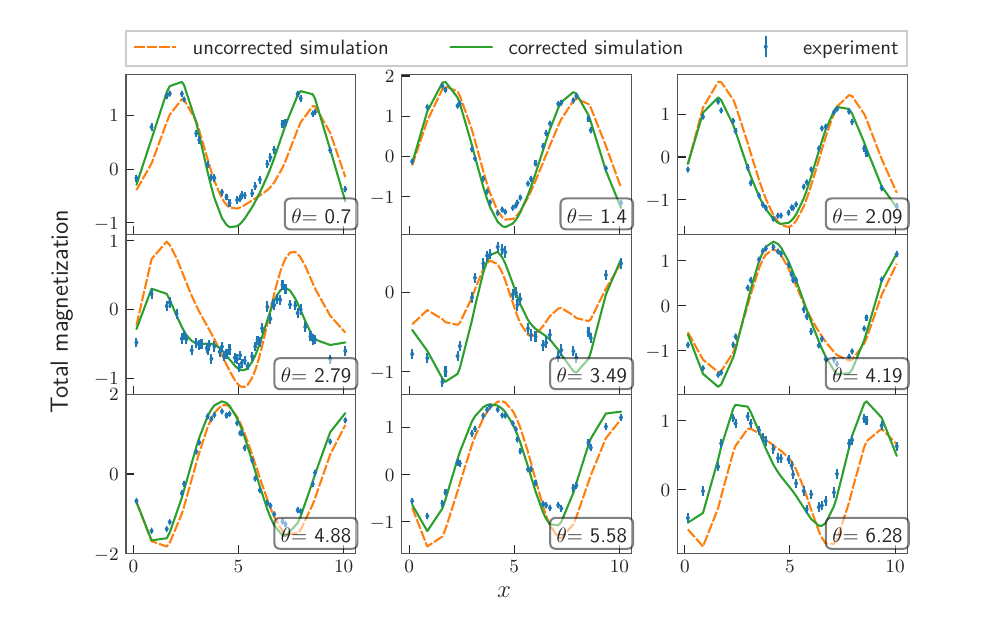}
    \caption{\emph{Data and simulation.} Data obtained from the \gls{naqpu} (blue dots) for 9 values of the ansatz parameter $\theta$ (legended for each plot), compared against \texttt{pulser} noiseless simulations. Error bars represent the Poissonian shot noise expected from the attained number of shots at each point. 
    The orange line is the result of the simulation using a default hardware configuration, whereas the green line is obtained by including the contribution from a detuning offset $\delta_{\text{offset}}$, as explained in the text.
    The x-axis represents the value of the input variable $x$, whilst the y-axis represents the total magnetization of the output, as defined in Eq.~\eqref{eq:circ_output}.
    \label{fig:data}}
\end{figure*}

We report in Fig.~\ref{fig:data} the results of the numerical simulations for the total magnetization -obtained as in Eq.~\eqref{eq:circ_output} using \texttt{Pulser}~\cite{pulser} - at the different feature $x$ values adopted for the experiment, mentioned in the main text, Sect.~\ref{sec:setup}. 
Though without any correction (orange line) the experimental results (blue dots) generally follow the expected trend, some deviation is observed. 

We identified the discrepancy to be explainable via an experimental offset altering the desired detuning $\delta$ value during the entire sequence. 
The Pulser simulation and subsequent analysis were corrected to account for this deviation, by including a $\delta_{\text{offset}}$ parameter in the \emph{EOM mode} of the emulator. 
The green line in Fig.~\ref{fig:data} corresponds to this ``corrected'' simulation, after optimizing $\delta_{\text{offset}} = \qty[parse-numbers=false]{-2{\pi}\times 162}{\kilo \hertz}$  to minimize RMSD weighted by statistical error while comparing the output data to the simulation. 
According to the hardware specifications, the $\delta$ chosen for the sequence is expected to be accurate up to $\sim \qty[parse-numbers=false]{2{\pi}\times 100}{\kilo \hertz}$, with the compensation for the light shift, which depends on the amplitudes that are used, introducing an additional uncompensated phase. Considering this, the value found for $\delta_{\text{offset}}$ is compatible with expectations.
Once this correction is applied, the experimental results match the simulation significantly better, as immediately evident observing the green line in Fig.~\ref{fig:data}.

\end{document}

%% file: glossary.tex
\glsdisablehyper

\newacronym{naqpu}{NA-QPU}{Neutral Atom - Quantum Processing Unit}
\newacronym{qpu}{QPU}{Quantum Processing Unit}
\newacronym{ftqc}{FTQC}{Falt-Tolerant Quantum Computer}

\newacronym{de}{DE}{Differential Equation}

\newacronym{nn}{NN}{neural network}
\newacronym{pinn}{PINN}{physics-informed neural network}
\newacronym{fm}{FM}{feature map}
\newacronym{ml}{ML}{machine learning}
\newacronym{gml}{GML}{graph machine learning}
\newacronym{sciml}{SciML}{scientific machine learning}
\newacronym{piml}{PIML}{Physics-informed machine learning}
\newacronym{ufa}{UFA}{universal function approximator}
\newacronym{wl}{WL}{Weisfeiler-Lehman}

\newacronym{qc}{QC}{Quantum Computing}
\newacronym{na}{NA}{Neutral Atoms}
\newacronym{nisq}{NISQ}{noisy intermediate scale quantum}
\newacronym{hea}{HEA}{Hardware Efficient Ansatz}

\newacronym{qsciml}{QSciML}{Quantum scientific machine learning}
\newacronym{qml}{QML}{quantum machine learning}
\newacronym{dqc}{DQC}{differentiable quantum circuits}
\newacronym{pqc}{PQC}{parameterised quantum circuit}
\newacronym{qcl}{QCL}{quantum circuit learning}
\newacronym{qnn}{QNN}{quantum neural network}
\newacronym{psr}{PSR}{Parameter Shift Rule}
\newacronym{agpsr}{aGPSR}{approximate Generalized Parameter Shift Rule}
\newacronym{qel}{QEL}{Quantum Extremal Learning}
\newacronym{qrc}{QRC}{quantum reservoir computing}